\documentclass{article}

\usepackage[utf8]{inputenc}
\usepackage[T1]{fontenc}
\usepackage[english]{babel}

\usepackage[letterpaper,top=2cm,bottom=2cm,left=3cm,right=3cm,marginparwidth=1.75cm]{geometry}
\usepackage{amsmath}
\usepackage{amsfonts}
\usepackage{amssymb}
\usepackage{mathtools}
\DeclarePairedDelimiterX{\infdivx}[2]{(}{)}{%
  #1\;\delimsize\|\;#2%
}

\usepackage{graphicx}
\usepackage[numbers,square,comma,sort&compress]{natbib}
\usepackage[colorlinks=true, allcolors=blue]{hyperref}
\makeatletter
\AtBeginDocument{
	\hypersetup{
		pdftitle = {\@title},
		pdfauthor = {D. N. Blaschke, T. Nguyen, M. Nitol, D. O'Malley, S. Fensin}
	}
}
\makeatother
\usepackage{subfig,epstopdf}
\title{\texorpdfstring{\begin{flushright}
			{\small LA-UR-23-20468}
		\end{flushright}\vspace{2em}}{}%
Machine Learning Based Approach to Predict Ductile Damage Model Parameters for Polycrystalline Metals}
\author{D. N. Blaschke$^1$, T. Nguyen$^2$, M. Nitol$^3$, D. O'Malley$^4$, S. Fensin$^3$
\\
$^1$XCP-5, Los Alamos National Laboratory\\
$^2$T-3, Los Alamos National Laboratory\\
$^3$MPA-CINT, Los Alamos National Laboratory\\
$^4$EES-16, Los Alamos National Laboratory\\
}
\date{June 6, 2023,\\[0.9ex]
\normalsize
Email: dblaschke@lanl.gov, thao.nguyen.me@lanl.gov, mash@lanl.gov, omalled@lanl.gov, saryuj@lanl.gov
}

\begin{document}
\maketitle

\begin{abstract}
Damage models for ductile materials typically need to be parameterized, often with the appropriate parameters changing for a given material depending on the loading conditions. This can make parameterizing these models computationally expensive, since an inverse problem must be solved for each loading condition. Using standard inverse modeling techniques typically requires hundreds or thousands of high-fidelity computer simulations to estimate the optimal parameters. Additionally, the time of a human expert is required to set up the inverse model. Machine learning has recently emerged as an alternative approach to inverse modeling in these settings, where the machine learning model is trained in an offline manner and new parameters can be quickly generated on the fly, after training is complete. This work utilizes such a workflow to enable the rapid parameterization of a ductile damage model called TEPLA with a machine learning inverse model.  The machine learning model can efficiently estimate the model parameters much faster, as compared to previously employed methods, such as Bayesian calibration.
The results demonstrate good accuracy on a synthetic test dataset and is validated against experimental data.
\end{abstract}

\section{Introduction}
Damage and failure in ductile materials is a complex function of microstructure and loading conditions.  Previous work has shown that microstructural features such as grain size, texture, and dislocation density can all affect spall strength by altering the number of void nucleation sites.  Additionally, for a given microstructure altering the peak velocity, tensile strain rate, and pulse duration (loading conditions) can all modify the amount of damage in a ductile material. 
Currently, ductile damage in materials is represented using various models such as TEPLA, TONKS, MNAG etc. \cite{addessio1993rate,Segurado:2012,nguyen2022TeplaManual,Tonks:2002,tonks2003,chen2021modified}.
However, these models need to be parameterized using experimental velocity-time and porosity-position data.  As the loading conditions are altered, the models need to be re-parameterized.  This not only makes the parameterization of models time consuming but also highlights the non-predictive nature of these models.  To overcome some of these challenges, we propose to use a machine learning model that can predict the parameters of a TEPLA model rapidly given specific loading conditions and material type.

Machine learning is becoming an increasingly important part of material science \cite{butler2018machine}.
Examples of applications include:
enabling rapid prototyping of new materials on a computer rather than in a laboratory \cite{jain2013commentary}, advances in batteries \cite{tran2011alloy}, identification of perovskite compositions for solar cells \cite{im2019identifying}, among many others. In data poor settings, scientists often must identify key features that can make learning easier for the model. However, in scenarios with relatively large datasets, deep learning algorithms can do this feature identification automatically \cite{lecun2015deep}. Some work more closely related to this manuscript investigated fracture dynamics in materials \cite{srinivasan2018quantifying}. Machine learning has been used to accelerate high-strain models of brittle fracture at the continuum scale by training a machine learning surrogate of a mesoscale model, which is used to inform a continuum scale model \cite{fernandez2021accelerating}.

Inverse modeling \cite{nakamura2015inverse} and the closely related field of inverse design \cite{sanchez2018inverse} are another fertile area for machine learning. The goal of inverse modeling is to determine the properties of a system by observing the behavior of the system. Typically, these are properties that are impractical to measure directly. Machine learning has been used to infer material properties based on, e.g., information about fluid flows in porous media \cite{kadeethum2021framework} or waves propagating through the medium \cite{wu2019inversionnet}. In the context of inverse design, machine learning can help design materials with desirable properties \cite{sanchez2018inverse}.

In this work, we aim to use inverse modeling to identify the appropriate parameters in a constitutive damage model. The physical setup for our model replicates a flyer plate experiment \cite{meyers1994dynamic,kanel2010spall}. The inputs to the physical model consists of information regarding the loading as described by velocity-time data (e.g., the flyer plate velocity) and the constitutive model parameters. The outputs of the physical model consist of the resulting porosity in the target plate after impact.

Machine learning models generally outperform traditional inverse modeling approaches because the machine learning model can run in milliseconds (with extra computations to generate the training data done in an offline mode). By contrast, traditional inverse methods run the physical model repeatedly in an iterative loop with each physical model run taking significant CPU resources that are typically on order of CPU-hours. This general trend holds true in our work and more details are given in the manuscript. Our approach enables one to rapidly go from an experimental result to a calibrated physical model. This enables modelers to rapidly begin modeling the behavior of the material after performing an experiment that characterizes the material without having to go through the process of setting up and running a traditional inverse model. It is worth noting that surrogate-based optimization approaches \cite{forrester2009recent} are a potential alternative to our machine learning inverse approach that would also provide significant computational acceleration compared to traditional inverse modeling approaches.

The rest of the manuscript is organized as follows.  Section~\ref{Methods} describes the details of the constitutive damage model, the data generation and the machine learning model.  Section~\ref{Results & Discussion} presents the results from the machine learning model and discusses the implications.  Finally, conclusions are presented in Section \ref{Conclusions}.

\section{Methodology}
\label{Methods}
This section discusses the details of the ductile damage model, TEPLA, data generation for training and the machine learning model used in this work.

\subsection{Brief description of TEPLA}
\label{sec: tepla}
\begin{figure}
    \centering
    \includegraphics[width=\textwidth]{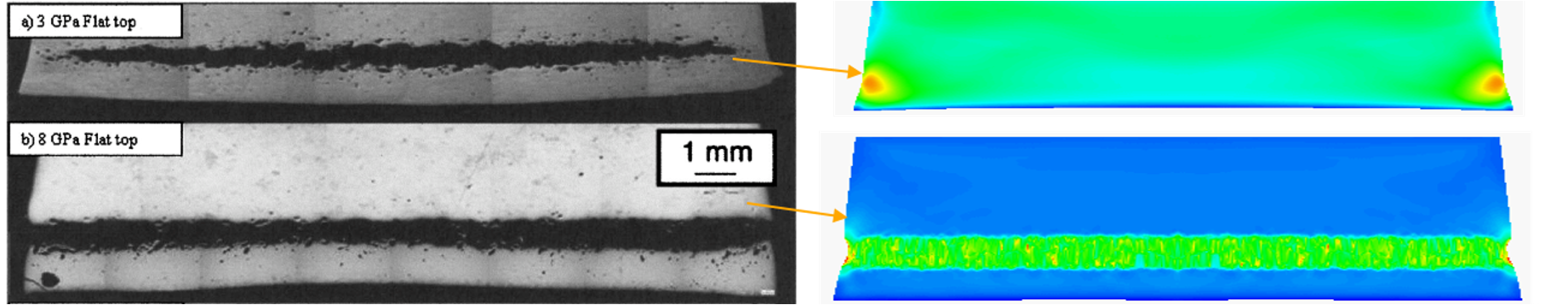}
    \caption{Figure showing (left) an example of optical micrographs highlighting varying amounts of damage in a metal (tantalum) as a function of peak stress, (right) simulation images showing predicted damage from a TEPLA model.  The green regions highlight areas of low density or high porosity.
    See \cite{Bronkhorst:2016} for related work on analyzing damage in tantalum.}
    \label{fig:my_label}
\end{figure}

The TEPLA model has been demonstrated as a model that works well in representing ductile damage and failure in materials especially under complex stress states.  Hence, it was used as an example model in this work.
\autoref{fig:my_label} illustrates an example of how damage can be measured and compared to a simulation using the TEPLA damage model.
Owing to the strong coupling between hydrostatic and equivalent stress components of the stress tensor in evolving damage, the TEPLA model has the capability to predict ductile damage in complex loading scenarios.
Therefore, the  TEPLA model is chosen in this work to illustrate the construction of a machine learning model to predict model parameters.

The original version of the TEPLA model was proposed by 
\cite{addessio1993rate}, based on the Gurson yield function \cite{gurson1977} for Cauchy stress tensor ($\boldsymbol{\sigma}_\text{eq}$), porosity (or void volume fraction, $\phi$) at the equilibrium state (denoted as subscript $_\text{eq}$), i.e.,
\begin{equation}
\label{eq: gurson equation}
F\left(    \boldsymbol{\sigma}_\text{eq};  \phi_\text{eq},T \right) 
= \left(\frac{\tau_\text{eq} }{ Y}\right)^2 
+ 2q_1 \phi_\text{eq} \cosh{\left[-q_2 \frac{3}{2} \frac{P_\text{eq}   }{ Y_\text{S} }\right]} 
-1 -q_3 \phi_\text{eq}^2.
\end{equation}
In the above equation, $Y$ is the flow stress of the fully dense material surrounding a void. Typically for application to the dynamic response of ductile metals, $Y$ is a function of equivalent plastic strain ($\bar{\varepsilon}_\text{m}^\text{pl}$), strain rate ($\dot{ \bar{\varepsilon}}_\text{m}^\text{pl}$), temperature ($T$), and density ($\rho_m$) (or pressure) of the material, i.e. $Y = \hat{Y}(\bar{\varepsilon}_\text{m}^\text{pl}, \dot{ \bar{\varepsilon}}_\text{m}^\text{pl}, T, \rho)$.
$Y_\text{S}$ is the saturation value of the flow stress (i.e. fully hardened) in the material surrounding the void, and here we approximate this behavior by assuming $Y_\text{S}$ is a constant. The hydrostatic pressure, $P$, corresponds to the macroscopic stress state, i.e. $\boldsymbol{\sigma}$, $P = -\frac{1}{3} \boldsymbol{\sigma} :\textbf{I}$, with \textbf{I} denoting the second order identity tensor. 
The macroscopic effective deviatoric stress (or von Mises stress) $\tau$ is calculated from the macroscopic deviatoric stress tensor ($\textbf{s} = \boldsymbol{\sigma}  +P \textbf{I}$) as $\tau = \sqrt{\frac{3}{2} \textbf{s}:\textbf{s} }$. 
The three parameters 
$q_1$, $q_2$ and $q_3$ were introduced by \cite{tvergaard1981influence} to account for the interaction of microvoids.
Here, we use the commonly accepted values of the three parameters, i.e.\ $q_1 = 1.5$, $q_2=1.0$, and $q_3=q_1^2$ originally suggested by \citep{tvergaard1981influence}.
The porosity ($\phi$) is related to the mass density of the fully dense matrix, $\rho_m$, and the macroscopic porous mass density, $\rho$, as $\phi=\frac{\rho_m}{\rho} - 1$. 
The initial value of porosity is denoted as $\phi_0$ (i.e.\ $\phi_0=\frac{\rho_m|_{t=0}}{\rho|_{t=0}} - 1$).
The failure value of porosity $\phi$, beyond which the material has no inherent resistance to macroscopic stress is denoted as $\phi_\text{f}$.

This viscoplastic model allows the possibility of the actual state of stress and porosity to exceed the flow threshold surface in \autoref{eq: gurson equation}. 
In the current version of TEPLA,
the viscoplastic flow is governed by an overstress relaxation kinetic expression akin to Duvaut-Lions viscoelasticity theory, i.e.,
\begin{equation}
\label{eq: duvaut-lions tensor}
\textbf{D}_\text{in} = \boldsymbol{\Pi}^{-1} (\boldsymbol{\sigma} - \boldsymbol{\sigma}_\text{eq}).
\end{equation}
In the above equation, $\boldsymbol{\Pi}$ is a fourth order viscosity tensor defined by its inversion, i.e.,
\begin{equation}
\label{eq: viscosity tensor}
\boldsymbol{\Pi}^{-1} = \left[{
\frac{1}{3\tau_\text{r}} 
\delta_\text{ij}\delta_\text{kl}  
}\right]
\textbf{e}_\text{i}
\otimes
\textbf{e}_\text{j}
\otimes
\textbf{e}_\text{k}
\otimes
\textbf{e}_\text{l}.
\end{equation}
The relaxation time, $\tau_\text{r}$, is defined as 
\begin{equation}
\label{eq: tau_r equation}
   \tau_\text{r} =  {\eta}\frac{1-\phi}{\phi},  
\end{equation}
in which $\eta$ is a material viscosity parameter and $\phi$ is the porosity.
The overstress relaxation expression in the TEPLA model (original and current versions) is enabled by ignoring the micro-inertia part in the model of \cite{johnson1981dynamic}.
Therefore, this simplification limits the use of the TEPLA model to scenarios where the micro-inertia effect is small.
A traditional rate-independent associated flow is recovered as the actual state approaches an equilibrium state on the threshold surface \citep{zuo2008implicit}. 
The definition of the viscoplastic flow of the current TEPLA model (\autoref{eq: duvaut-lions tensor}) is modified from the original TEPLA model \cite{addessio1993rate}, in order to completely separate the viscosity due to void growth and the underlying shear strength model of the solid material. This separation enables the current TEPLA model to better capture both the damage signal from the velocity history in plate impact tests (which is affected by the underlying shear strength model) and the porosity distribution (which is affected by only the viscosity due to void growth) of the same simulation, without compromising the mesh sensitivity of the TEPLA model. 
In addition, the relaxation time $\tau_r$ of the current TEPLA model (\autoref{eq: tau_r equation}) is also modified from the original TEPLA model. 
\autoref{eq: tau_r equation} has proven to be more effective in reproducing features of the pullback signal in velocity time history observed in flyer plate impact experiments.
The current TEPLA model used in this work was implemented into FLAG, an arbitrary Lagrangian-Eulerian (ALE) multiphysics code developed at Los Alamos National Laboratory \citep{caramana1998construction}.
Further details on the current TEPLA model and its numerical implementation was presented in \cite{nguyen2022TeplaManual}.

With the three parameters $q_1$, $q_2$ and $q_3$ specified, the remaining parameters to be calibrated for each polycrystalline metal are $Y_S$, $\phi_0$, $\phi_f$, and $\eta$. It is worth noting that porosity at failure ($\phi_f$) is only sensitive to measurements where failure happens, such as shock and spall loading.
These remaining three parameters ($Y_S$, $\phi_0$, and $\eta$) have a significant coupled effect on representing ductile damage in simulations (such as velocity-time history observed and the porosity distribution). Therefore, multiple measurements (such as a combination of both velocity history and porosity distribution from one loading condition or multiple velocity histories from several loading conditions) are needed to better constrain the parameters. Additionally, every time the material or loading conditions are altered, the model usually needs to be re-parameterized.
This can be an expensive and slow process which thus motivates our present machine learning approach.

\subsection{Data generation}
\label{sub: data generation}

\begin{table}[ht]
\centering
\begin{tabular}{r|ccc|cc}
	& $Y_\text{spall}$ [MBar] & $\log\phi_0$ & $\eta$ [MBar\,$\mu$s] & $v_\text{impact}$ [cm/$\mu$s] & flyer thickness [cm] \\\hline
copper, min& 0.0004 & -3.8 & 0.00027 & 0.015 & 0.1 \\
max & 0.005 & -2.4 & 0.001 & 0.05 & 0.2 \\\hline
aluminum, min & 0.0004 & -3.8 & 0.000055 & 0.04 & 0.03 \\
max &  0.005 & -2.4 & 0.001 & 0.08 & 0.08 \\\hline\hline
copper, Fig.~\ref{fig:exampletraining},
A & 0.000647 & -2.6828 & 0.000652 & 0.019051 & 0.192960\\
B & 0.004546 & -2.5084 & 0.000478 & 0.018357 & 0.100973\\
C & 0.002028 & -2.6396 & 0.000531 & 0.019625 & 0.123273\\
D & 0.002172 & -2.4131 & 0.000324 & 0.018920 & 0.163291\\
E & 0.004926 & -2.9052 & 0.000335 & 0.021633 & 0.119036\\\hline
\end{tabular}
\caption{We show the ranges of the five parameters we varied to generate the training data.
Additionally, the lower part of the table lists the (randomly chosen) sets of parameters used in \autoref{fig:exampletraining} below.}
\label{tab:trainingdataranges}
\end{table}

The four parameters in the TEPLA model $Y_\text{S}$, $\phi_0$, $\phi_\text{f}$, and $\eta$, are usually calibrated to measured free surface velocity time histories and porosity distribution (when data are available) from flyer plate impact experiments.
Both velocity and porosity measurements are known to be sensitive to the chosen parameters for the damage models \cite{thissell1damage, nguyen2021bayesian}. 
In the following, we will provide a brief introduction of the plate impact experiments and how the measured velocity-time history and final porosity distribution are affected by the parameters of the damage model.

A detailed description of plate impact experiments can be found elsewhere \citep{meyers1994dynamic,kanel2010spall}. One of the important features in these experiments is the interaction of two rarefaction waves, which leads to a strong tensile pulse leading to  ductile failure of the material through nucleation, growth, and coalescence of voids. The free surface velocity is measured as a function of time using  photon Doppler velocimetry. Information regarding the details of the loading conditions can be obtained from the velocity-time data.  In general, as the porosity increases, the impedance of the material changes, which creates a pullback signal in the velocity-time data.

Within the acoustic limit, the velocity difference at pullback and the maximum velocity (at the Hugoniot state) can be linearly related to the tensile strength of a material \citep{kanel2010spall}. 
In particular, the quasistatic part of the tensile strength is proportional to $\frac{2}{3}Y_S \ln(1/\phi_0)$ \citep{carroll1972static}; the viscosity part of the tensile strength is proportional to the material viscosity \cite{cortes1992dynamic}, i.e.\ $\eta$ in the TEPLA model. 
In addition, the porosity distribution in the recovered sample indirectly reflects the material tensile strength. 
The porosity distribution can be characterized by the peak value of porosity and its spans. 
A higher material tensile strength (with a lower value of $\phi_0$, or higher values of $Y_S$ and/or $\eta$) results in a lower peak porosity and narrower span. 
As a failure criterion, the porosity at failure ($\phi_f$) only affects the velocity signal and porosity distribution once it is satisfied. In the porosity distribution, the consequence of having porosity larger than $\phi_f$ is the discontinuity of porosity since the shear resistance reduces to zero and the porosity could grow without resistance according to the applied volumetric strain. In the velocity signal, the rise from the pullback signal is sharp and the next local maximum velocity is almost as high as the velocity plateau at the Hugoniot state. By definition, the porosity at failure ($\phi_f$) is not sensitive to experiment measurements where incipient spall is observed. In the scenarios of having only incipient spall data, calibration of $\phi_f$ results in only a lower limit for its values rather than a finite value. Therefore, we adopt a reasonable value of $\phi_f=0.25$ from \cite{tvergaard1981influence} in those scenarios.   

The velocity signal and the associated porosity distribution within the material are useful in calibrating a ductile damage model such as TEPLA.
In most experiments, only the velocity history data are available while the corresponding porosity distribution is not.  This is because porosity measurements require a special experimental apparatus to soft-recover the specimen without causing further significant damage.

Ideally, the TEPLA model should be calibrated in a wide range of shock pressures and \textit{(tensile)} strain rates, for each material of interest. This requires the availability of data from plate impact experiments in the corresponding shock pressures and strain rates. However, such diverse data are usually not available for every single material due to the high cost of performing well-controlled experiments. 
Therefore, the TEPLA model needs to be re-calibrated when the material changes, or when the loading condition is faraway  from the calibrated loading condition range.

To determine the dependencies of TEPLA model parameters on the loading conditions (velocity-time data), FLAG simulations were performed using a baseline TEPLA model.  These synthetic data were also used as input data for the machine learning model used in this study. 
Specifically, flyer plate impact experiments were modeled at varying peak velocities and pulse durations using a 1D array of 3D computational zones along the thickness of the flyer plate and specimen. In particular, the flyer and target plates were meshed with 200 zones per 1.5 mm along the thickness, respectively.
This mesh size is fine enough to adequately dampen the ringing of the simulated velocity at the Hugoniot state (i.e.\ the flat plateau of the velocity profile, see \cite{forquin2017pulse}, \autoref{fig:predicted_accuracy}b, blue line, for an example) together with the artificial bulk viscosity. 
The ringing is the result of discretizing the continuous solid in which a shock wave is propagating in.
In addition, the mesh size was found to provide a converged solution for both the velocity and porosity distributions during the tensile loading phase, with our current specimen materials. 

\begin{figure}
\centering
\includegraphics[width=0.49\textwidth]{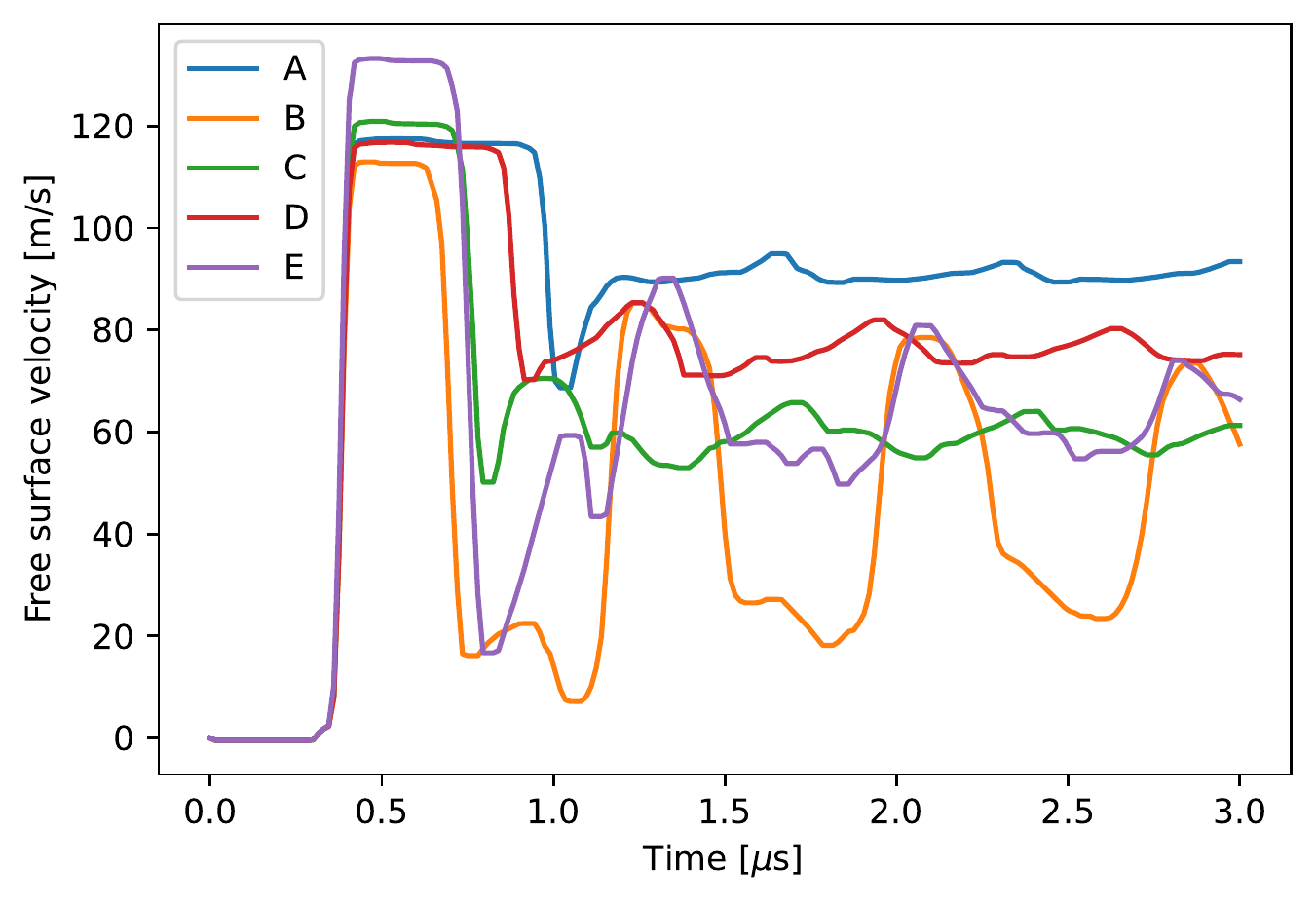}%
\includegraphics[width=0.51\textwidth]{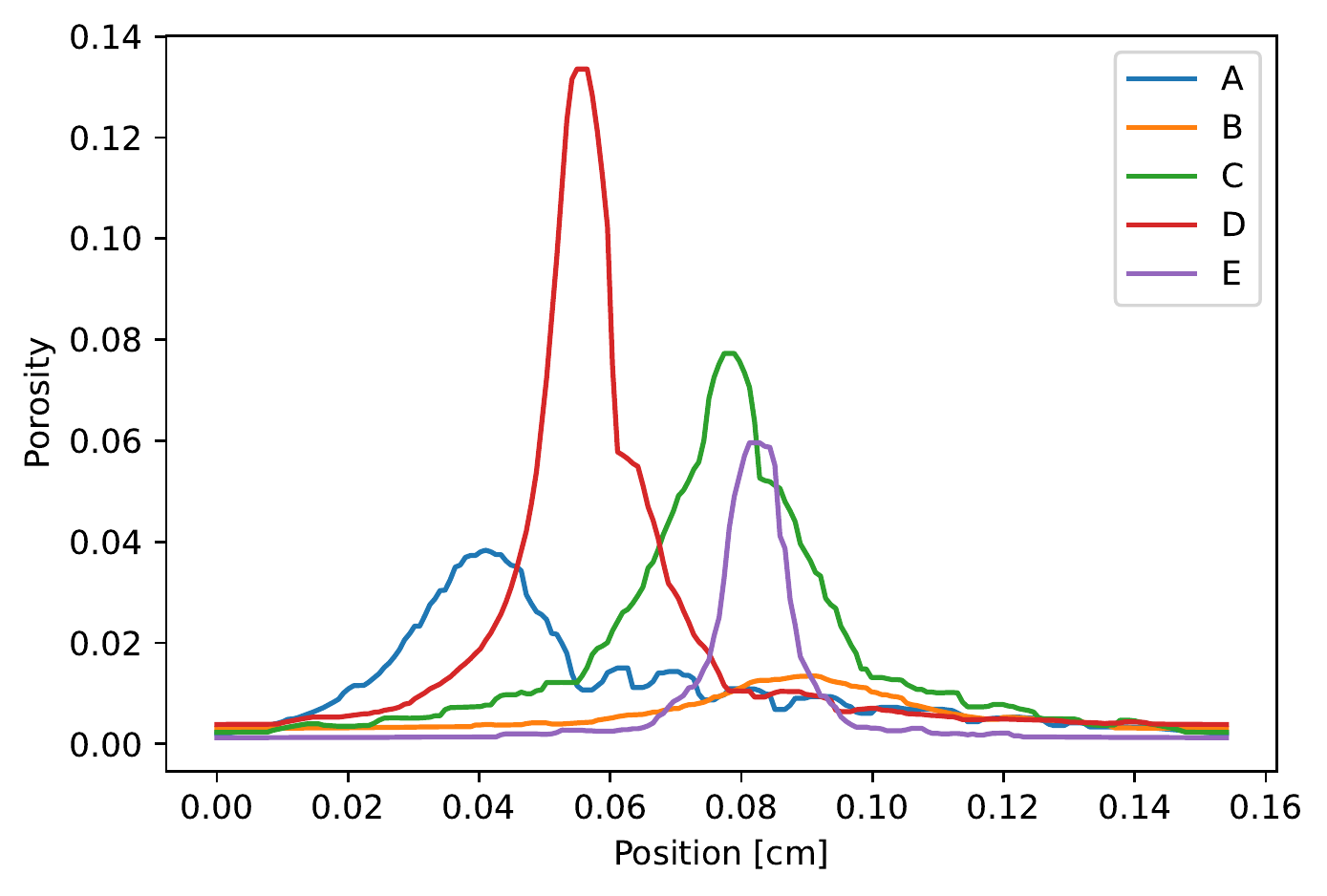}
\caption{Examples of training data for half hard copper showing changes in the pulse duration and peak stress during loading, i.e. the free surface velocity-time (left) and the corresponding porosity vs. position (right) as a function of random changes to TEPLA parameters, impact velocity, and flyer thickness.
The five sets of (randomly chosen) parameters used to simulate these data are listed in \autoref{tab:trainingdataranges}.
The data shown here were smoothed using a median filter.}
\label{fig:exampletraining}
\end{figure}

For each material studied here, we performed 3000 simulations with different TEPLA parameters, impact velocities, and flyer thicknesses (to vary the pulse duration).
All five parameters were varied simultaneously using the Latin hypercube sampling method.
The ranges of those parameters are assembled in \autoref{tab:trainingdataranges}.
3000 simulations required about 1100 core hours of computation time.
For each loading condition and the corresponding final porosity distribution, the velocity profiles resulting from this set of simulations (or simulation observations) were then extracted and used as training data for the machine learning model.
\autoref{fig:exampletraining} illustrates some examples of training data.

The job of the machine learning model was to analyze the nonlinear relationship between the input parameters and the simulation observations, and to provide the solution of the inverse problem when experimental observations were provided. 
Training the neural network (for 2000 epochs), as described in the following subsection, was fast and took less than 20 minutes on a quad-core laptop with data from 3000 simulations.

\subsection{Machine learning}
In this application of machine learning, a deep learning neural network (NN) for processing arrays of synthetic input, was employed. The architecture of a neural network was a multi-layered feed-forward neural network created by successively stacking numerous hidden layers. This deep architecture enables neural networks to learn key features needed to make accurate predictions. In our networks, the hidden layers consisted of dense layers followed by activation layers, and in certain cases, pooling layers. 

Here, the NN input training data comprise of porosity and velocity data that are output by the FLAG hydro code simulations as described above. The goal of the machine learning model is to predict the TEPLA model parameters $Y_{\mathrm{S}}$, $\phi_{0}$, and $\eta$.  The train-test split method was used to evaluate the performance of an algorithm for machine learning. In general, this assessment may be used for any supervised learning technique. As part of this technique, a dataset is divided into two subsets: The first subset, known as the training dataset, is utilized to fit the model and the second subset is not used to train the model; rather, the input element of the dataset is delivered to the model, after which predictions are generated and compared to the expected values. This second batch of data is usually known as the test data. 

The performance of  machine learning algorithms is enhanced when numerical input variables are scaled to a standard range. We use this technique in our application of machine learning. Standardization is used to scale numerical data prior to modeling in this instance. Standardization adjusts the distribution to have a mean of zero and a standard deviation of one by rescaling each input variable independently by removing the mean and dividing by the standard deviation. This is how the standardization was implemented: $z=\frac{x-\mu}{\sigma}$ where $\mu$ is the mean and $\sigma$ is the standard deviation. The mean and variance of each characteristic in our data are computed, and then all features were transformed using their respective mean and variance.

\begin{figure*}[!htbp]
\centering{
\captionsetup[subfigure]{justification=centering}
\subfloat[ ]{\includegraphics[width=0.5\textwidth]{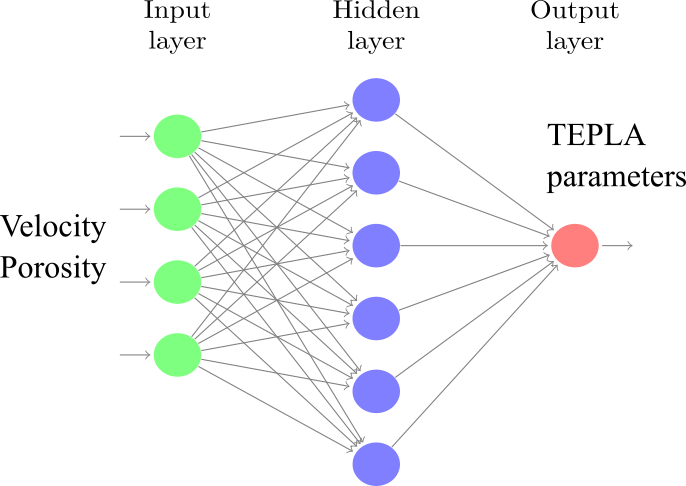}}
\subfloat[ ]{\includegraphics[width=0.5\textwidth]{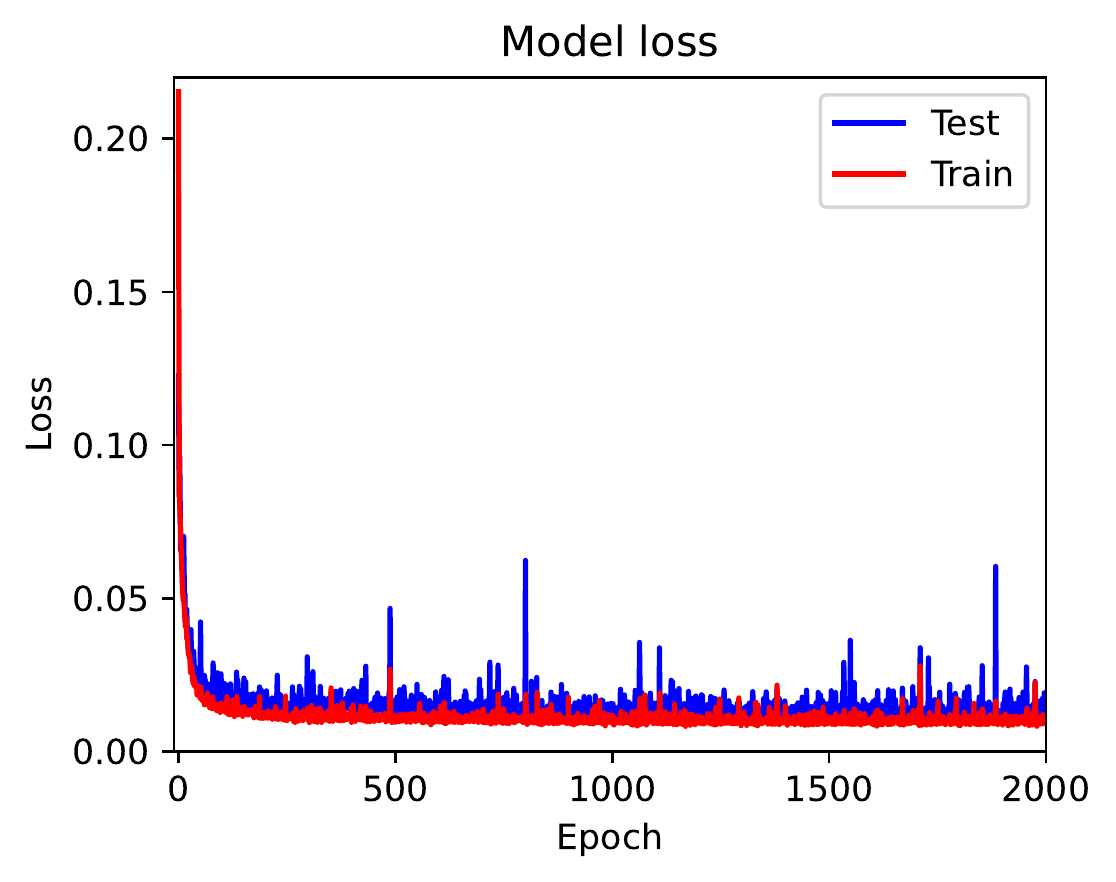}}}
\caption{(a) Schematic showing the various layers of the neural network along with input and output data (b) The variation in the loss function for the training and test datasets as the model is trained}
\label{model}
\end{figure*}

Following the scaling of the data, the ML model was built using Keras \cite{chollet2015keras}, a high-level Python API for constructing and training deep learning models. Keras is versatile, including the ability to create complicated models. The completely linked layers are defined using the Dense class, with the first parameter specifying the number of neurons in the layer and the activation function specifying the activation argument. The first hidden layer contain 50 nodes and is activated by the \emph{LeakyRelu} function. The conventional \emph{ReLU} function is almost equivalent to the \emph{LeakyRelu} function. The \emph{LeakyRelu} trades hard-zero sparsity for a gradient that may be more resistant during optimization. Unlike the normal \emph{ReLU} function, the \emph{LeakyRelu} has a non-zero gradient across its whole domain. The purpose for utilizing the \emph{LeakyRelu} activation function is the negative component is completely dropped in \emph{ReLU}, whereas it has a non-zero slope in \emph{LeakyRelu}. The \emph{LeakyRelu} is capable of retaining some of the negative values that flow into it. This increased output range provides the model with more flexibility. In \emph{LeakyRelu}, incorporating a nonzero slope for a negative portion enhances the results. Regularization is used to avoid overfitting, which adds a penalty as model complexity increases. The regularization parameter penalizes all parameters except the intercept, ensuring that the model generalizes the data and does not overfit. In this case, a regularizer with $alpha$ parameters 1e-5 and 1e-4 is employed to apply both L1 and L2 regularization penalties. In L1 regularization or Lasso regression, the coefficients are determined by minimizing the L1 loss function, which is denoted as follows: $J\left(w\right)_{L1} = \sum\limits_{i=1}^{n}\left( y^{\left(i\right)} - \hat{y}^{\left(i\right)}\right)^{2}+\alpha \sum\limits_{j=1}^m\left| w_j\right|$ and in L2 regularization or Ridge regression, the coefficients are determined by minimizing the L2 loss function, given as: $J\left(w\right)_{L2} = \sum\limits_{i=1}^{n}\left( y^{\left(i\right)} - \hat{y}^{\left(i\right)}\right)^{2}+ \alpha\sum\limits_{j=1}^{m} w_{j}^{2}$.  In elastic net regularization, the combination of L1 and L2 regularizers is utilized. Regularization penalties are added to the loss function during training in this extension of linear regression. 
The regularization hyperparameters were set to their default values and were not optimized. There are also hyperparameters associated with the number of neurons in the network. We did not attempt to optimize these hyperparameters and wouldn't expect significant changes in performance if we did. 
The advantage of elastic net is that it allows for a balance of both penalties, which can result in greater performance than a model with either one or the other penalty on the fitting in this problem. The second hidden layer is made up of 200 neurons and uses the same approach as the first hidden layer. The output layer uses a rectified linear unit activation function from the last layer. The mean squared error (MSE) is used as a regression loss function. The loss is calculated as the mean of the squared discrepancies between true and anticipated values. Because of the squaring element of the function, the MSE is useful for guaranteeing that the trained model does not have any outlier predictions with big errors. Instead of the traditional stochastic gradient descent process, the adaptive movement estimation algorithm, or Adam, is utilized to update network weights iteratively depending on training data. The Adam optimization approach is a stochastic gradient descent extension that has lately found increased usage for deep learning. It automatically adapts a learning rate for each input variable for the objective function and further smooths the search process by updating variables with an exponentially decreasing moving average of the gradient. \autoref{model} shows a (simplified) schematic of a neural network and 2000 iterations of fitting.

\section{Results \& Discussion}
\label{Results & Discussion}

\begin{figure}[!htb]
\centering
\includegraphics[width=\textwidth]{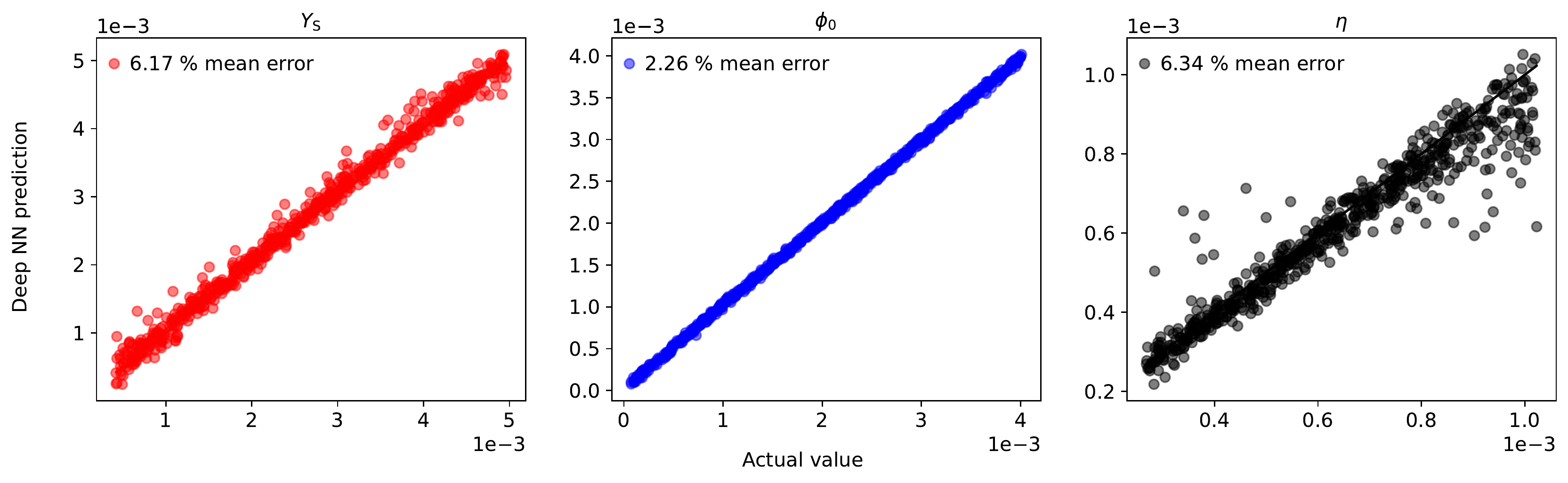}
\caption{Figure showing predicted vs. actual values of the TEPLA parameters for half hard copper.
We used a median filter on the training data which consisted of porosity over position data, free surface velocity as a function of time, as well as impact velocity and impact pules duration (resp. flyer plate thickness).
The units of $Y_S$ and $\eta$ are MBar and MBar$\cdot\mu$s, respectively. The parameter $\phi_0$ is dimensionless.
}
\label{fig:predicted_accuracy}
\end{figure}

\begin{figure}[!htb]
\centering
\includegraphics[width=\textwidth]{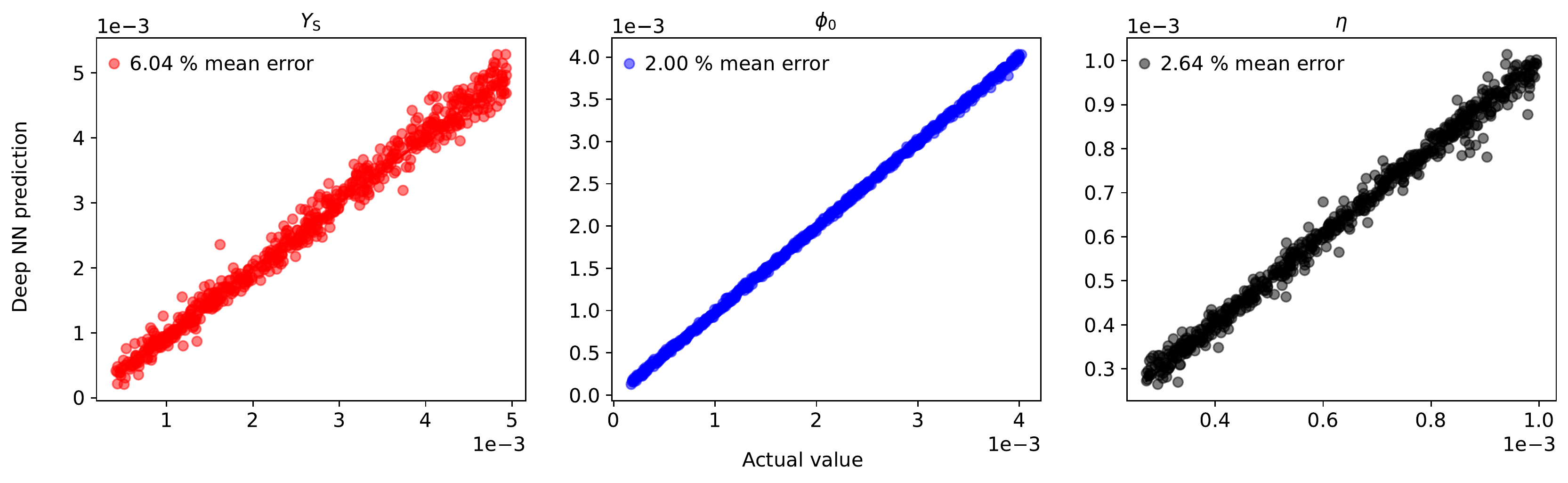}
\caption{Figure showing predicted vs. actual values of the TEPLA parameters for annealed copper.
We used a median filter on the training data which consisted of porosity over position data, free surface velocity as a function of time, as well as impact velocity and impact pules duration (resp. flyer plate thickness).
The units of $Y_S$ and $\eta$ are MBar and MBar$\cdot\mu$s, respectively. The parameter $\phi_0$ is dimensionless.
}
\label{fig:predicted_accuracy2}
\end{figure}

\begin{figure}[!htb]
\centering
\includegraphics[width=\textwidth]{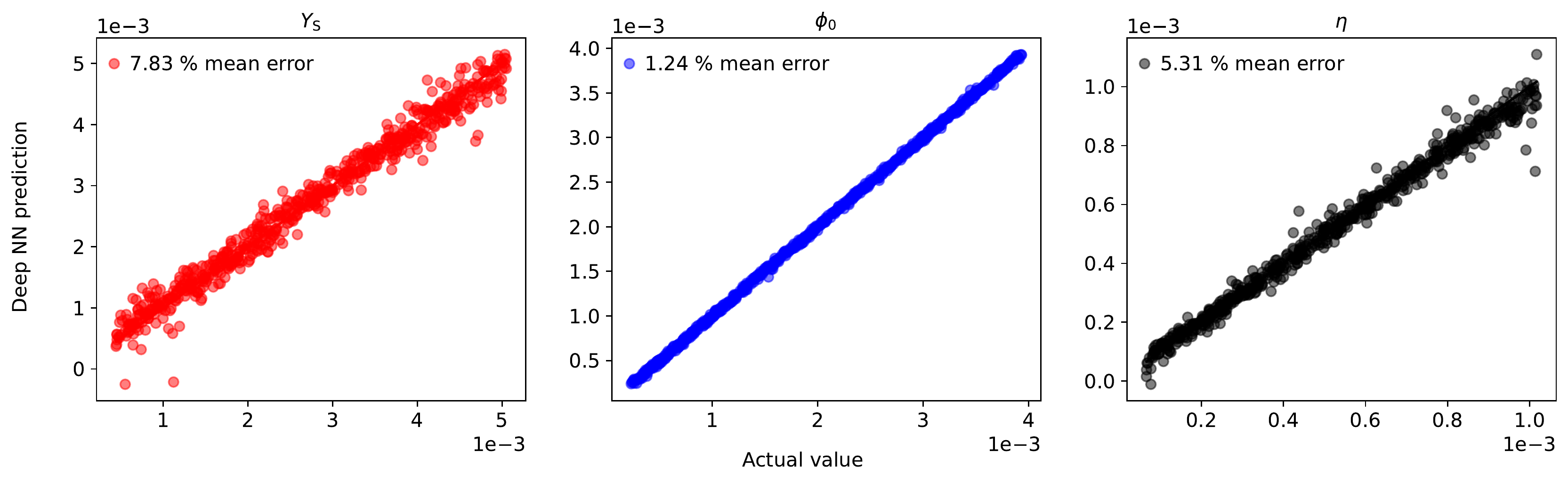}
\caption{Figure showing predicted vs. actual values of the TEPLA parameters for aluminum 6061.
We used a median filter on the training data which consisted of porosity over position data, free surface velocity as a function of time, as well as impact velocity and impact pules duration (resp. flyer plate thickness).
The units of $Y_S$ and $\eta$ are MBar and MBar$\cdot\mu$s, respectively. The parameter $\phi_0$ is dimensionless.
}
\label{fig:predicted_accuracy3}
\end{figure}

\begin{figure}[!htb]
\centering
\includegraphics[width=0.48\textwidth]{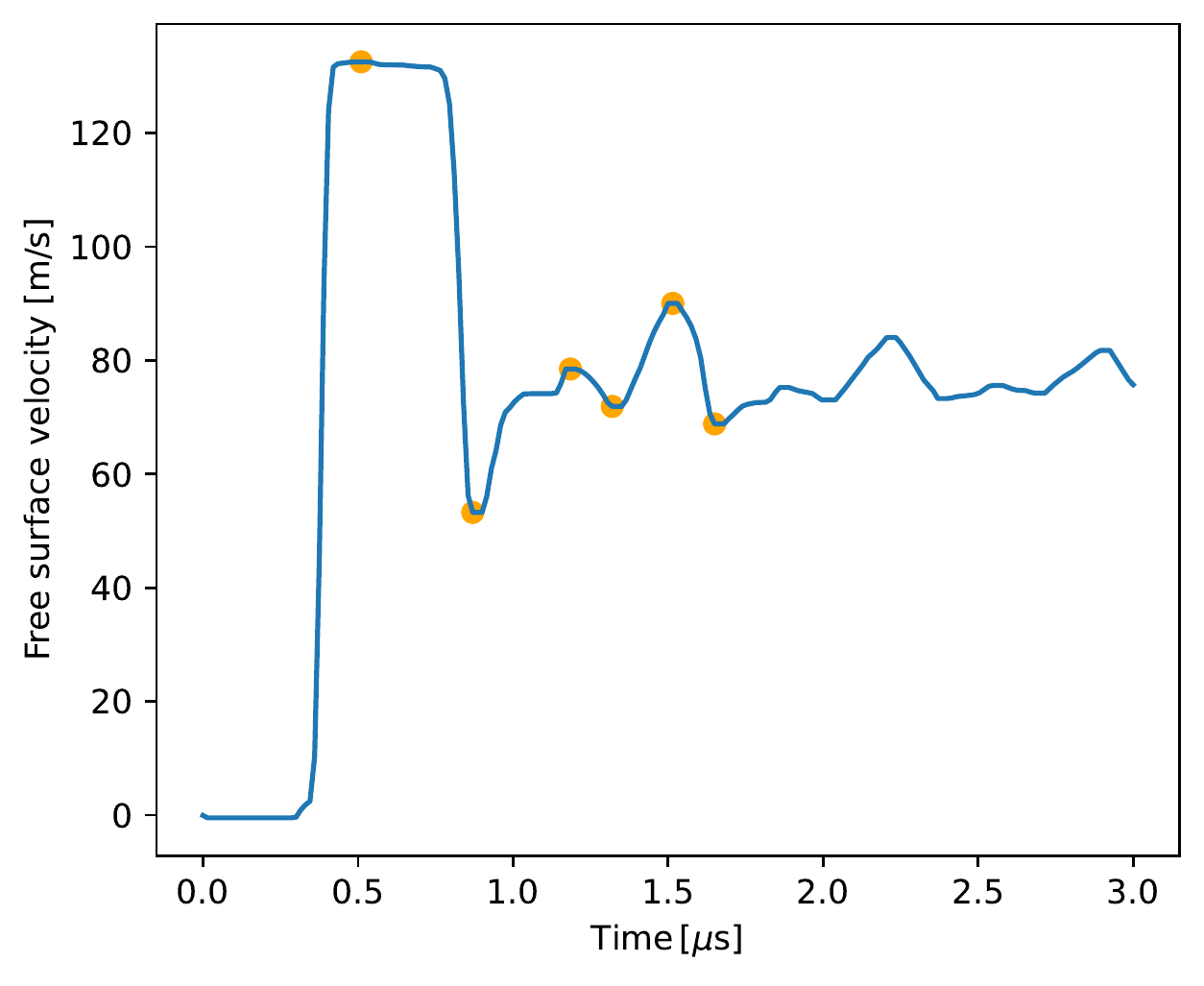}%
\includegraphics[width=0.49\textwidth]{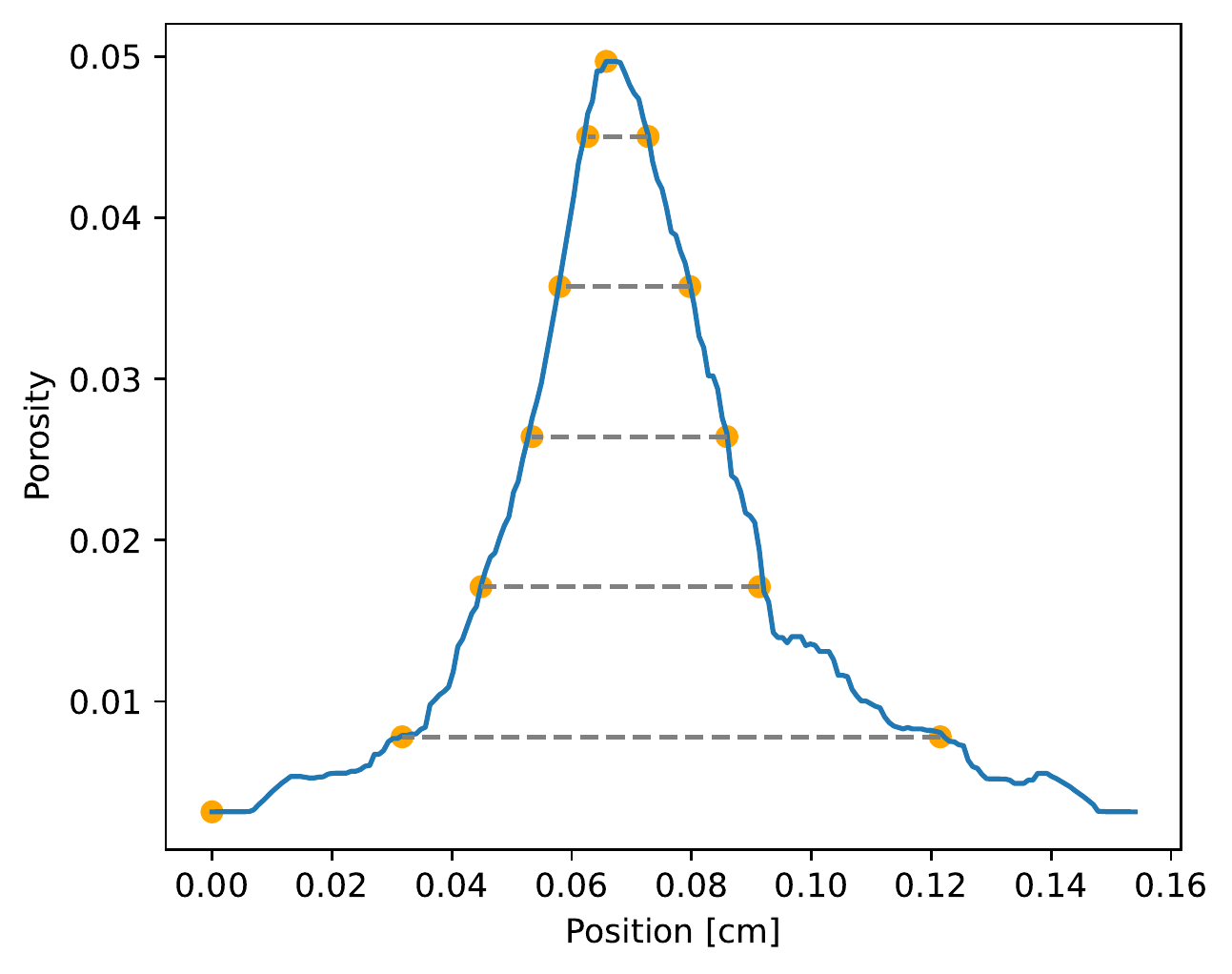}
\caption{We highlight which `features' are extracted from a typical simulation result for the purpose of training our ML model.
From the free surface velocity (left subfigure), only the first three maxima and subsequent minima (not their times stamps) are stored.
From the porosity data (right subfigure) we extract the porosity values of the global minimum and maximum, as well as the widths (i.e. lengths of the gray dashed lines) of where the porosity exceeds 10\%, 30\%, 50\%, 70\%, and 90\% of the maximum minus the minimum.
}
\label{fig:featureextraction}
\end{figure}

\begin{figure}[!htb]
\centering
\includegraphics[width=\textwidth]{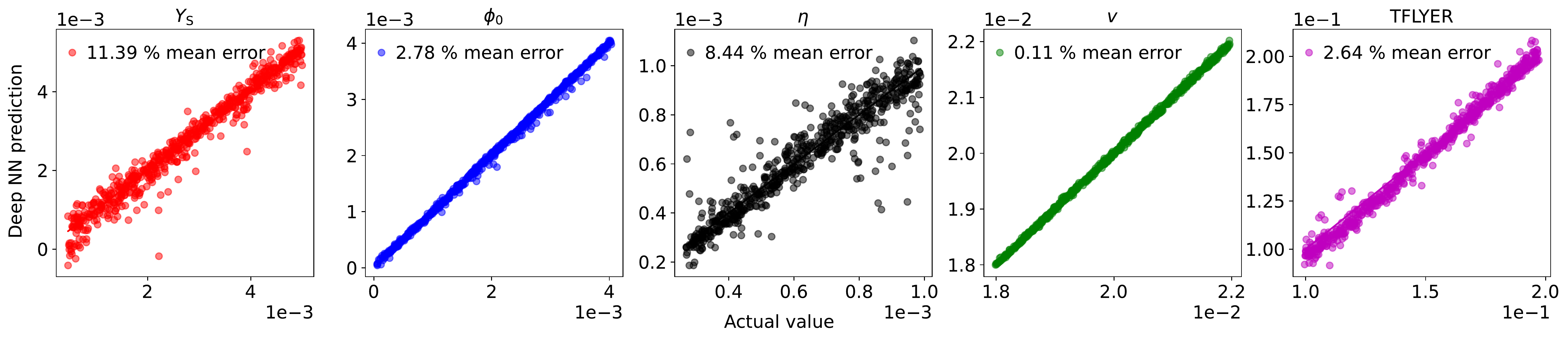}
\caption{Predicted versus actual values of the TEPLA parameters for half hard copper.
In this example, we used as training data only a limited number of `features' extracted from porosity over position data and free surface velocity as a function of time after applying a median filter, as described in the main text.
In addition to the three TEPLA parameters, we have trained the neural network in this example to additionally predict the impact velocity $v$ [cm/$\mu$s] and the thickness of the flyer (TFLYER [cm]).
Units are the same as in \autoref{tab:trainingdataranges}.
}
\label{fig:predicted_accuracy4}
\end{figure}

\begin{figure}[!htb]
    \centering
    \includegraphics[width=0.5\textwidth]{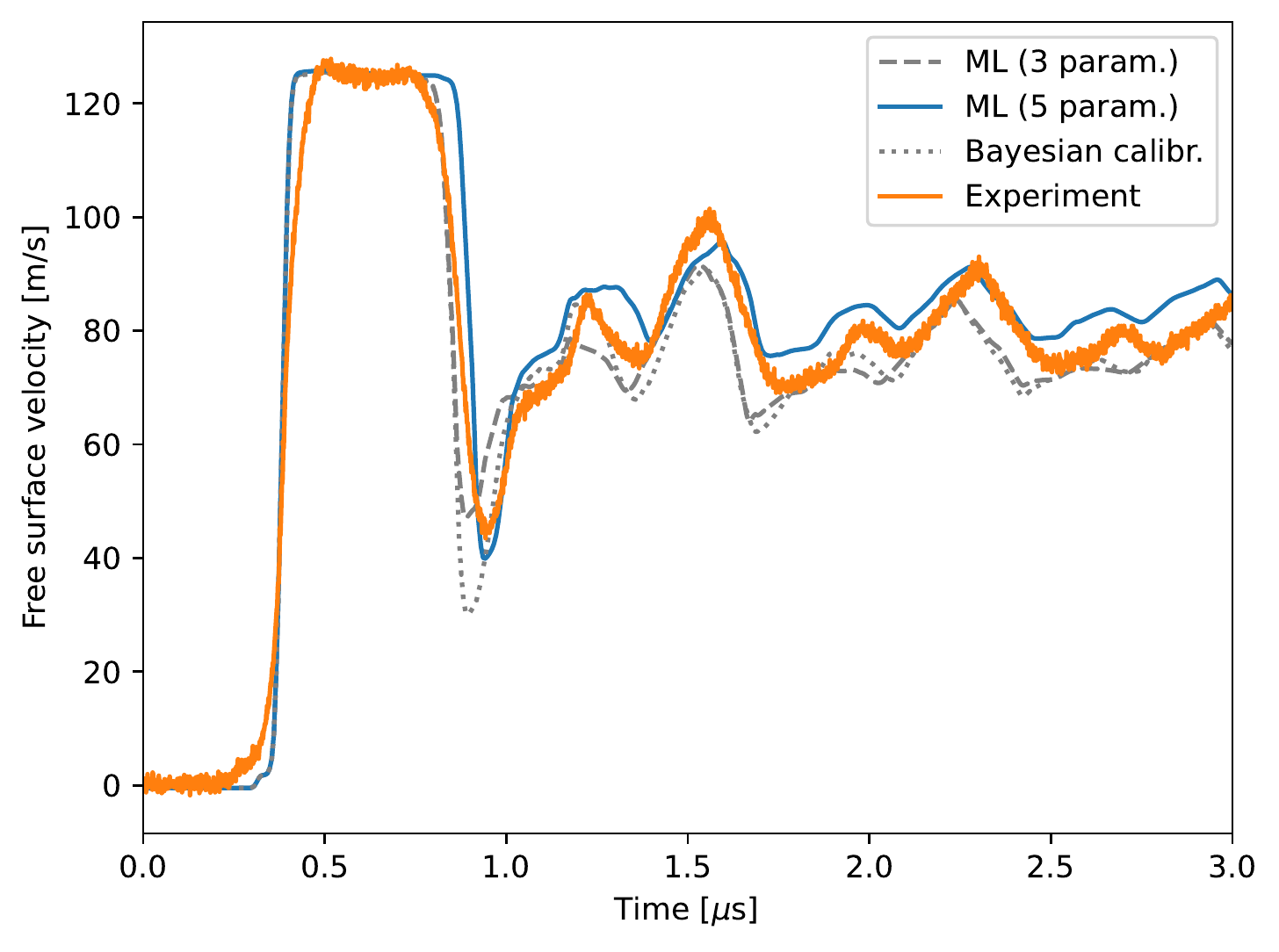}%
    \includegraphics[width=0.5\textwidth]{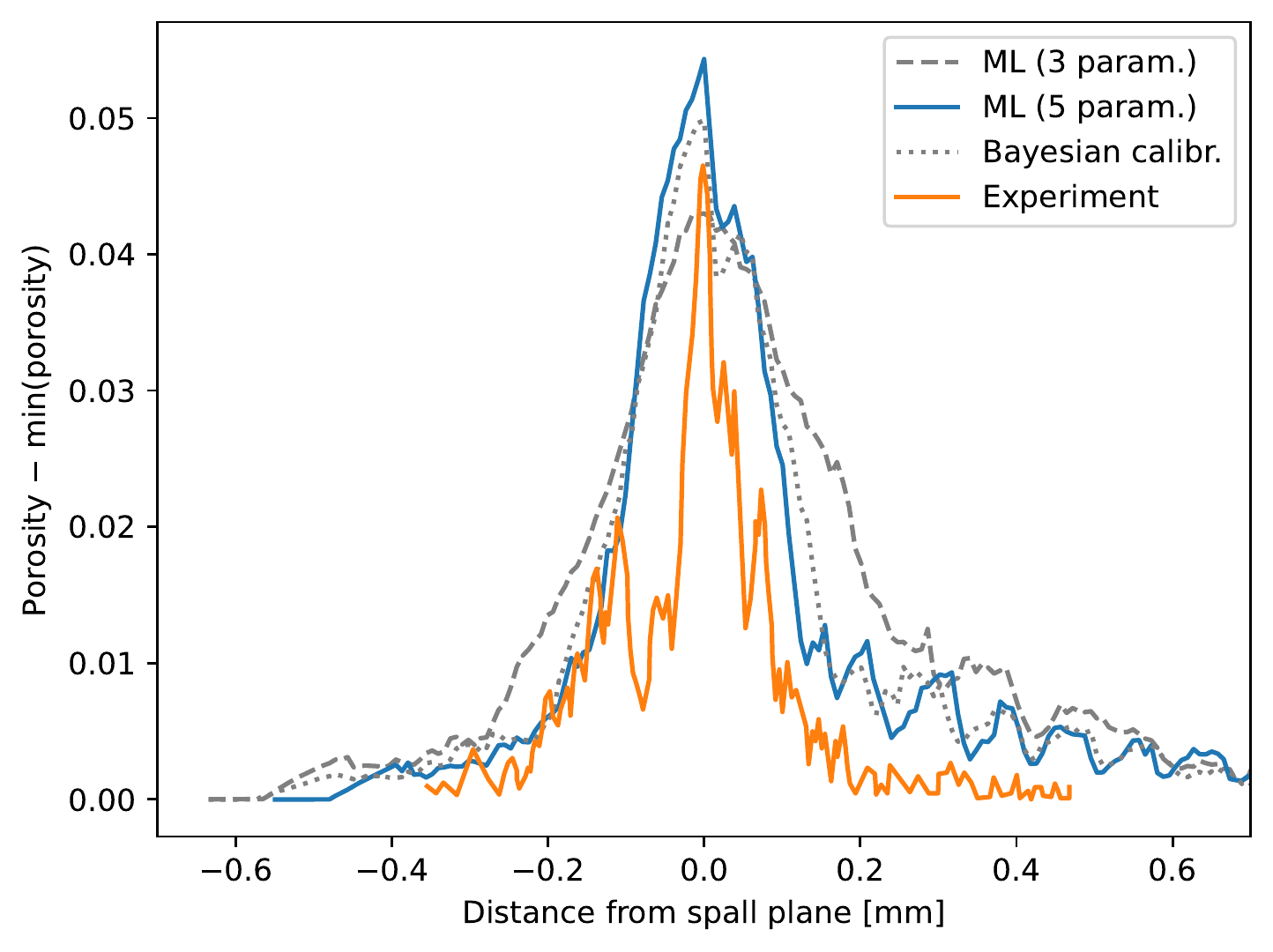}
    \caption{Comparison between predicted and experimentally measured velocity and porosity profiles for half hard copper.
    The resolution in position for the copper target within the FEM simulation was $\Delta x=7.7\,\mu$m.
    The labels 'ML (3 param.)' and 'ML (5 param)' refer to neural networks trained (using the reduced dataset) to predict only the 3 TEPLA parameters or to predict all 5 varied parameters (i.e. 3 TEPLA parameters, flyer thickness, and impact velocity).
}
    \label{fig:compare_velocities_porosities}
\end{figure}

The performance of our machine learning model is demonstrated by simulating plate impact experiments for three materials: half-hard copper, annealed copper, and aluminum 6061.
These simulations used equations of state described in Refs. \cite{Peterson:2012} (copper) and \cite{Crockett:2004} (aluminum).
Furthermore, a PTW strength model \cite{PTW:2003} as well as the TEPLA damage model were used in this study \cite{addessio1993rate,thissell1damage, nguyen2021bayesian,nguyen2022TeplaManual}.
The PTW model is a shear strength model suitable for plastic deformation of polycrystalline metals at high strain rates. 
In this model, the yield strength and saturated shear strength is scaled with the material shear modulus, which is a function of temperature and pressure. 
The rate dependence of the flow
stress is calculated using Wallace’s theory for overdriven shocks in metals \cite{Wallace:1981a,Wallace:1981b}.
The PTW model parameters for aluminum are taken from \cite{plohr2022ptw}.
The PTW model parameters for half-hard copper are modified from \cite{prime2016using}, in order to match the yield strength implied by free surface velocity at Hugioniot state. The PTW model parameters for copper are presented in \autoref{tab:ptw half-hard copper parameter}.

\begin{table}[ht!]
\centering
\begin{tabular*}{\columnwidth}{@{\extracolsep{\stretch{1}}}lr|r|r|rr@{}} 
& \textbf{Parameters}& \textbf{Unit} & \textbf{Half-hard copper}& \textbf{Annealed copper}&\\\hline
& $p$ & $-$ & 1.0 & 2.23& \\
& $\theta$ & $-$ & 0.0057& 0.0198 & \\
& $\kappa$ & $-$ & 0.0926 &  0.1887
 &\\
& $\gamma$ & $-$ & 0.000007 & 0.000063&\\
& $s_0$ & $-$ & 0.007033 & 0.009427&\\
& $s_{\infty}$ & $-$ & 0.000226$^*$& 0.001081& \\
& $y_0$ & $-$ & 0.000363$^*$ & 0.000303 &\\
& $y_{\infty}$ & $-$ & 0.000224$^*$ & 0.000134&\\
& $y_1$ & $-$ & 0.057 & 0.094& \\
& $y_2$ & $-$ & 0.8 & 0.575& \\
& $\beta$ & $-$ & 0.25 & 0.25 &\\\hline
\end{tabular*}
\caption{
\label{tab:ptw half-hard copper parameter}
PTW model parameter values for half-hard copper and annealed copper used in this work.
Parameters for half-hard copper marked with $^*$ are modified from the work of \citep{prime2016using}. 
 }
\end{table}

For each one of the three materials, three thousand plate impact scenarios were simulated with different impact velocities, flyer thicknesses (to simulate different pulse duration during shock), as well as the TEPLA model parameters $\phi_0$, $\eta$, and $Y_\text{S}$ which are described above in Subsection \ref{sec: tepla}; ranges are given in \autoref{tab:trainingdataranges}.
The resulting porosity as a function of position within the target as well as the free surface velocity as a function of time were then used to train an ML model using the methodology described above.
In order to reduce noise in the porosity and velocity data, a median filter with a median filter window size of 5 was employed (i.e. scipy's signal.medfilt() function with kernel\_size=5).
The flyer thickness and the impact velocity were also considered input data to train the ML model to predict the TEPLA model parameters.
20\% of the FLAG simulations were retained as test data, and
Figs.~\ref{fig:predicted_accuracy}, \ref{fig:predicted_accuracy2}, and \ref{fig:predicted_accuracy3} highlight the accuracy of the ML model on those test data for the three different materials studied here.

It is noted here that the initial porosity $\phi_0$ yields a higher prediction accuracy than the two remaining parameters in these cross-validation results. This is because our training data here contain the porosity at the impact and free ends of the target plate (see the left graph on \autoref{fig:exampletraining}). 
At these two ends, the material is subjected to compression rather than tension (which is the essential condition for porosity growth in the TEPLA model) during the course of loading. 
As a consequence, the porosity is unchanged (i.e. $\phi=\phi_0$) at these two ends. Using the unchanged porosity (i.e.\ $\phi_0$) in the training data would bring the high accuracy in porosity prediction, shown in Figs.~\ref{fig:predicted_accuracy}, \ref{fig:predicted_accuracy2}, and \ref{fig:predicted_accuracy3}.
On the other hand, parameters $Y_S$ and $\eta$ are only inferred indirectly based on the velocity profile and porosity distribution, leading to a lower level of cross-validation accuracy (especially in half hard copper, where we observe a small number of outliers in the $\eta$ test data of \autoref{fig:predicted_accuracy}).

Although in general both flyer thickness and impact velocity are known for each experiment, these numbers can also be easily predicted from the surface velocity and porosity data, as demonstrated in \autoref{fig:predicted_accuracy4}.
Furthermore, the current ML model can achieve almost comparable accuracy when the training data are focused to reproduce a limited set of `features' extracted from the porosity and surface velocity data. This approach ensures that the ML model is robust in predicting TEPLA parameters for experimental data which contain ranges and resolutions of porosity data that differ from the training data, for example.
The surface velocity features extracted were the first three maxima and subsequent minima (after filtering out `noise' using a median filter).
As for the porosity data, both the maximum and minimum value were extracted as well as the width of the porosity distribution containing porosities greater than 10\%, 30\%, 50\%, 70\%, and 90\% of the maximum.

These features represent the porosity distribution after the impact test. 
As mentioned in the third paragraph of Section \ref{sub: data generation}, these porosity and velocity features are sensitive to the material tensile strength, or the four parameters $Y_\text{S}$, $\phi_0$, $\phi_\text{f}$, and $\eta$ of the TEPLA model. \autoref{fig:featureextraction} shows an example of which features are extracted from the results of one simulation taken from the training dataset.
\autoref{fig:predicted_accuracy4} showcases how accurate a ML model trained with such a `reduced' dataset can be at the example of half-hard copper (compare with \autoref{fig:predicted_accuracy}).

In \autoref{fig:compare_velocities_porosities} we show a comparison between results from FLAG simulations and ML predicted TEPLA parameters using the reduced training data described above to an actual experimental dataset for half-hard copper taken from Ref. \cite{thissell1damage} (orange curve).
The predicted TEPLA parameters in this example (blue curves in \autoref{fig:compare_velocities_porosities}) were:
$Y_S\sim0.0022$ MBar, $\phi_0\sim0.00092$, $\eta = 0.00046$ MBar$\cdot\mu$s.
Furthermore, the ML model predicted the impact velocity at 204.2 m/s, which is very close to the true value of 203.4 m/s, as well as a slightly thicker flyer at 1.6 mm (instead of 1.5 mm).
Forcing the true values for flyer thickness and impact velocity (gray dashed curves in \autoref{fig:compare_velocities_porosities}), yields a similar prediction for the TEPLA parameters where the main difference is a slightly lower spall parameter of $Y_S\sim0.0021$ MBar.
The former prediction led to a slightly better match in the porosity profile.
The experimental data shown in orange were taken from \cite{thissell1damage}.

For comparison, we also show how the (much faster) ML technique compares to the Bayesian calibration technique (gray dotted curves in \autoref{fig:compare_velocities_porosities}) to determine TEPLA parameters. The Bayesian calibration procedure for TEPLA parameters follows the previously published work on ductile damage model calibration for single crystal plasticity \cite{nguyen2021bayesian}.
In this Bayesian calibration, the three TEPLA parameters $\phi_0$, $\eta$, and $Y_S$ were calibrated for half-hard copper against the same velocity profile and porosity distribution provided by \cite{thissell1damage}; while the impact velocity and flyer thickness are taken as their true values (i.e. 204.2 m/s and 1.6 mm, respectively). The posterior distribution of the Bayesian calibration process was constructed from Markov chain Monte Carlo sampling using $\sim70$ CPU-hours. 
The mean parameter values from the  posterior marginal distribution of the three calibrated parameters are then used as the set of Bayesian calibrated parameters.
Overall, the good agreement between the simulated velocity and porosity distribution from the TEPLA simulations and the corresponding experiment results implies a balance of the calibrated parameters. 
For example, just increasing the viscosity parameter $\eta$ leads to a wider spread of the  porosity distribution (i.e. higher standard deviation) and lowers the velocity slope from the pullback signal (i.e.\ acceleration due to re-compression).
On the other hand, a sole reduction of the initial porosity $\phi_0$ leads to a lower velocity at the pullback signal, and a smaller peak of porosity values across the target plate thickness.

\section{Conclusion}
\label{Conclusions}
This work illustrates the utility of machine learning inverse models in the context of ductile damage models. The machine learning model used here is able to rapidly predict a set of parameters ($Y_S$, $\phi_0$, and $\eta$ of the TEPLA damage model \cite{addessio1993rate,nguyen2022TeplaManual}) that enable accurate simulations of materials, such as half hard copper, annealed copper, and aluminum 6061, undergoing ductile damage. This machine learning model dramatically reduces the amount of computational time and model setup time required to go from an experiment of one of these materials to a parameterized ductile damage model, which is often a critical bottleneck in modeling these materials.

While trained on simulation data, the model was validated against both simulated and experimental data. Of course, the experimental data have some characteristics that differ from the simulation data. A key to ensuring the machine learning model remained accurate on the experimental data was identifying critical features in the data that could be used by the machine learning model. This also reduces the need for a very deep network and a very large dataset, as well as highlighting the importance of expert knowledge.

\subsection*{Acknowledgements}
We thank the referees for their valuable and insightful comments.
We also thank DJ Luscher for valuable feedback as well as C. Bronkhorst for help with Fig. \ref{fig:my_label}.

This work was supported by the Advanced Technology Development and Mitigation (ATDM) project within the Advanced Simulation and Computing (ASC) Program of the U.S. Department of Energy under contract 89233218CNA000001.

\subsection*{Data Availability}
Data are available at \url{https://doi.org/10.17632/r4hgs982vs.1}.

\bibliographystyle{utphys-custom}
\bibliography{sample}

\end{document}